\documentclass[11pt,preprint2]{aastex} 

\newcommand{\VLA}{\textit{VLA }}

\slugcomment{Accepted to the Astrophysical Journal}

\begin{document}

\title{Spectral Components of the Radio Composite Supernova Remnant
G11.2$-$0.3}

\author{Cindy Tam, Mallory S. E. Roberts \altaffilmark{1,2}, and
Victoria M. Kaspi \altaffilmark{2,3} }

\affil{Department of Physics, Ernest Rutherford Physics Building,
McGill University, 3600 University Street, Montreal, Quebec,
H3A 2T8, Canada}

\altaffiltext{1}{Quebec Merit Fellow}

\altaffiltext{2}{Department of Physics and Center for Space Research,
Massachusetts Institute of Technology, Cambridge, MA 02139}

\altaffiltext{3}{Alfred P. Sloan Research Fellow}

\begin{abstract}
We present a high-resolution radio study of the supernova remnant
(SNR) G11.2$-$0.3 using archival \VLA data.  Spectral
tomography is performed to determine the properties of this
composite-type SNR's individual components, which have only recently
been distinguished through X-ray observations.  Our results indicate
that the spectral index of the  pulsar wind nebula (PWN), or plerion,
is $\alpha_{P} = 0.25^{+0.05}_{-0.10}$.  We observe a spectral index
of $\alpha_{S} = 0.56 \pm 0.02$ throughout most of the SNR shell
region, but also detect a gradient in $\alpha$ in the south-eastern
component.  We compare the spectral index and flux density with recent
single-dish radio data of the source.  Also, the radio efficiency and
morphological properties of this PWN are found to be consistent with
results for other known PWN systems.
\end{abstract}

\keywords{pulsars: individual (PSR J1811$-$1925) --- supernovae:
individual (G11.2$-$0.3) --- supernova remnants --- radio continuum:
ISM}

\section{INTRODUCTION}
The proposed association between G11.2$-$0.3 and the supernova (SN)
event of 386~\textsc{ad} \citep{clar77} has prompted many years of
detailed study of this object.  Radio and X-ray observations made with
the \textit{Very Large Array} (\textit{VLA}) and \textit{Einstein}
satellite by \citet{down84} verified the existence of a highly
symmetric shell.  Downes then argued that G11.2$-$0.3 was likely the
result of either a 300-500 yr old Type I SN explosion like Tycho or
Kepler's supernova remnant (SNR), or an 
explosion similar to the event which produced SNR Cassiopeia A, but
which took place in 386 \textsc{ad}.  \citet{mors87} detected possible
flat spectrum emission in the central region from high-frequency radio
maps made with the \textit{Effelsberg} 100-m telescope, and concluded
that G11.2$-$0.3 belongs to a plerion-shell composite SNR class.
Overall spectral index estimates of $\alpha = 0.56$ \citep{down84} and 
$\alpha = 0.49$ \citep{mors87}, where $S_{\nu} \propto \nu^{-\alpha}$,
were determined from old and recent integrated flux density
measurements. Further \VLA observations \citep{gree88} produced high
resolution maps at 20 and 6 cm, and revealed clumpy emission at the
outer shell boundry, giving it the appearance of an evolved Type II
SNR similar to Cas A.  \citet{gree88} also estimated a distance of
$\sim 5$ kpc to the remnant based on its \ion{H}{1} spectrum, and a
diameter of $\sim 6$ pc.  \citet{reyn94} further supported the
hypothesis linking  G11.2$-$0.3 to the historical event through
correlations between radio and \textit{ROSAT} X-ray observations, but
argued for a Type Ia progenitor SN.
\par 
Hard-spectrum, non-thermal X-ray emission from a source within the
remnant was detected in data from \textit{ASCA}, which implied the
existence of an embedded pulsar producing plerionic emission
\citep{vasi96}, and confirmed that G11.2$-$0.3 is a composite remnant.
A 65-ms X-ray pulsar (PSR J1811$-$1925) was discovered by
\citet{tori97}; subsequent observations showed the characteristic
spin-down age and rate of rotational kinetic energy loss to be $\tau = 
24,000$ yr and $\dot E = 6.4 \times 10^{36}$ erg/s,
respectively \citep{tori99}.  The exact location of pulsar PSR
J1811$-$1925 and the X-ray morphology of the pulsar wind nebula (PWN)
were revealed for the first time in \textit{Chandra} X-ray images
\citep{kasp01}.
\par
Recent high-frequency radio observations using the single-dish
\textit{Effelsberg} telescope revealed the spectral index of
G11.2$-$0.3 to be $\alpha = 0.50$ for the combined spectrum of the PWN
and SNR \citep{koth01}.  A spectral index of $\alpha = 0.57$ for the
shell emission alone was also calculated, by subtracting an estimated
flat spectrum plerionic component.
\par
Because G11.2$-$0.3 has long been mis-classified as a purely
shell radio SNR, here we re-examine this source in light of its
composite nature, as recently revealed by hard X-ray observations.  In
this paper, we re-analyse archival \VLA data of G11.2$-$0.3 to
determine its spectral index, using the technique of spectral
tomography, which is a method that allows us to determine $\alpha$ for
both the shell and plerionic regions separately.  We
compare the flux density and spectral parameters of each feature
with results from single-dish radio studies.  We also measure the
structure and radio efficiency of the PWN to verify that they are
consistant with other known plerionic remnants.  These results
will also be included in interpreting X-ray data results (Roberts et
al., in preparation).

\section{OBSERVATIONS AND RESULTS} 
Radio observations of G11.2$-$0.3 were made with the \VLA at
20 and 6 cm (L- and C-bands, respectively), between 1984 April and
1985 May.  Details of these observations are summarized in
Table~\ref{tab:obs}. 
\begin{deluxetable}{lcccc}
\tablewidth{400pt}
\tablecaption{\VLA observational parameters for G11.2$-$0.3
\label{tab:obs} }
\tablehead{
\colhead{Observing} & \colhead{Array} & \colhead{Frequencies} &
\colhead{Bandwidth} & \colhead{Time on Source} \\
\colhead{Date} & \colhead{Configuration} & \colhead{(MHz)} &
\colhead{(MHz)} & \colhead{(min)} } 
\startdata
1984 Apr 26 & C   & 1465, 1515 & 50.00 & 39.5 \\
1984 Jun 06 & C   & 1465, 1635 & 50.00 & 18   \\
1984 Jun 06 & C   & 4535, 4965 & 50.00 & 19   \\
1984 Jul 05 & DnC & 1446, 1496 & 12.50 & 55.5 \\
1984 Jul 05 & DnC & 4923, 4873 & 25.00 & 144  \\
1984 Aug 27 & D   & 4535, 4965 & 50.00 & 14.5 \\
1984 Sep 09 & D   & 4835, 4885 & 50.00 & 34.5 \\ 
1985 Feb 22 & A   & 1415, 1490 & 6.25  & 77   \\
1985 Feb 25 & A   & 1415, 1490 & 6.25  & 82.2 \\
1985 May 19 & B   & 1415, 1490 & 6.25  & 62.3 \\
1985 May 19 & B   & 4640, 4670 & 6.25  & 129  \\
1985 May 25 & B   & 1415, 1490 & 6.25  & 63.3 \\
1985 May 25 & B   & 4640, 4670 & 6.25  & 112.8 \\
\enddata
\end{deluxetable}
\par
Data reduction and analysis were performed using standard
procedures within the \textsc{miriad} package \citep{saul99}.  The
data were flux-density and antenna-gain calibrated using the primary
calibrator 3C286 (J2000 1331+305).  The secondary calibrators
1749+096, 1748$-$253 and 1741$-$038 (B1950) were chosen for phase
calibrations, based on observation coordinates and wavelength band.
To aid in the process of detection and removal of bad visibilities, we
performed calibrations on each data set individually.  The header
files were adjusted to match pointing centers; this step was necessary
to prompt \textsc{miriad}'s restoring process into maintaining the
smallest possible synthesized beam.  
\par
\begin{figure}[!h]
\plotone{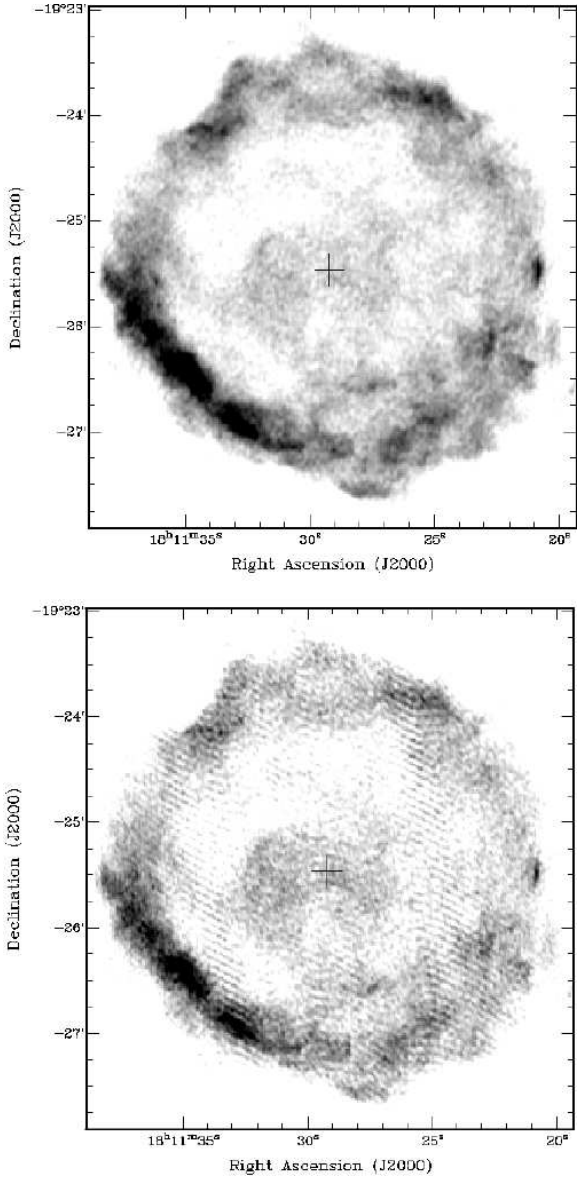}
\caption{High resolution images of SNR G11.2$-$0.3 at 20 (top) and 6
(bottom) cm.  The cross indicates the location of its associated
pulsar PSR J1811$-$1925.  Note that the size of the cross is much
larger than the positional uncertainty \citep{kasp01}.  The
``stripes'' in the 6-cm image are an artifact of emission from a
bright point source located to the south (not shown). \label{fig:snrs} } 
\end{figure}

Combined images at both 20 and 6 cm were formed using multi-frequency
synthesis and a uniform weighting scheme.  The images were deconvolved
using a maximum entropy algorithm \citep{nara86}.  We applied
iterative phase self-calibration procedures in multi-frequency mode to
make additional corrections to the antenna gains as a function of
time.  The resulting cleaned maps, as displayed in
Figure~\ref{fig:snrs}, were convolved with a Gaussian restoring beam
of dimensions $2\farcs6 \times 1\farcs8$ and $2\farcs1 \times 1\farcs5$
for 20 and 6 cm, respectively. 

\subsection{Spectral Index Determination}
To measure the spectral index $\alpha$, where $S_\nu \propto
\nu^{-\alpha}$, from the 20- and 6-cm images, we first needed to
spatially filter their $u-v$ sky distributions in order to match
spatial scales \citep{gaen99,craw01}.  We degraded the nearly-complete
$u-v$ coverage of the 20-cm image to that at 6 cm through a process of
$u-v$ modelling, which essentially created a 20-cm visibility dataset
with the same coverage as for the 6-cm image.  This was imaged and
deconvolved in the same way as were the original 6-cm data, and both
maps were convolved to 10$''$ resolution to aid clarity in the
following tomography process.
\par
We could then compare the two images in order to make a spectral index
determination using the method of spectral tomography \citep{katz97}.
A difference image $I_{\alpha_t}$ was calculated by scaling the 6-cm
image by a trial spectral index $\alpha_t$, and subtracting it from
the 20-cm image, according to the formula 
\[I_{\alpha_t}=I_{20}-\left(\frac{\nu_{20}}{\nu_6}\right)^{\alpha_t}I_6\ ,\]
where $I_{20}$ and $I_6$ were the images being compared, and
$\nu_{20}$ and $\nu_6$ were the average frequencies of the combined
20- and 6-cm images, respectively.  When the trial spectral index
reached the actual spectral index of a particular feature,
i.e. $\alpha_t = \alpha$, that region appeared to vanish into the
local background of the difference image.   Uncertainties in $\alpha$
were estimated by finding the range of $\alpha_t$ beyond which
significant residuals appeared.  If a certain area of the feature
possessed a slightly different spectral index, it appeared as a
distinctly positive or negative residual compared to the rest of the
difference image.

\subsection{Tomography Results}
By examining the series of difference images, we find approximate
values of $\alpha$ for the PWN and SNR regions of G11.2$-$0.3.
Positive residuals appear as light emission against the neutral grey
background, and negative as dark emission. 
\begin{figure}[!h]
\epsscale{0.8}
\plotone{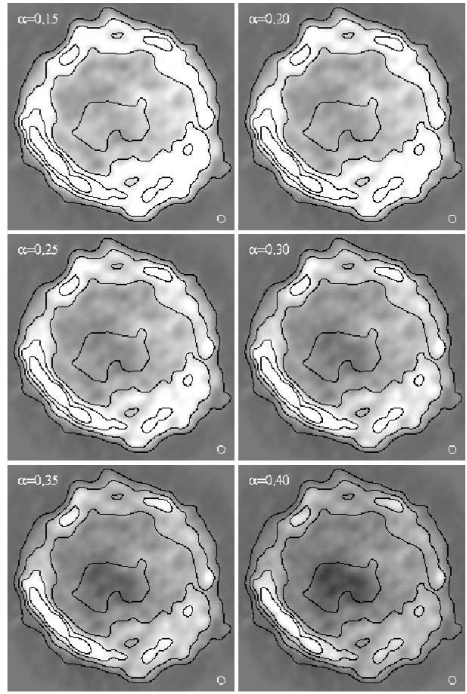}
\caption{Tomographic spectral index images of the PWN region of
G11.2$-$0.3.  The greyscale shows difference images for $\alpha_t =
0.15$ to $0.40$ in steps of 0.05.  Contours correspond to 6-cm data
convolved with a 10$''$ circular beam (shown in lower right corners),
at levels of 5\% to 41\% of peak flux 91 mJy/beam, in steps of
12\%. \label{fig:pwntomog} }
\end{figure}

\begin{figure}[!h]
\epsscale{0.8}
\plotone{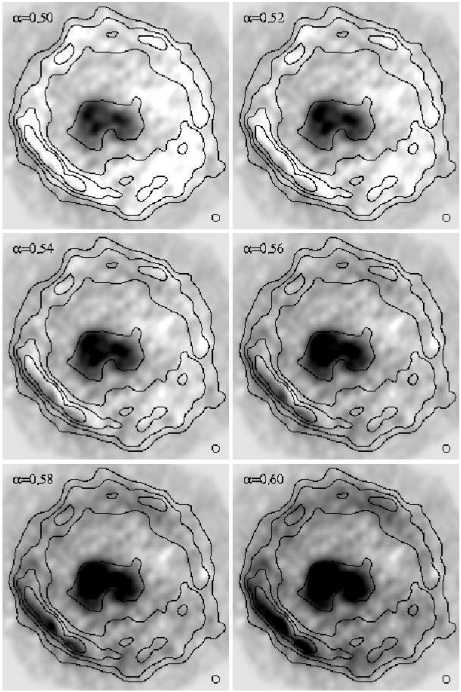}
\figcaption{Tomographic spectral index images of the SNR shell region
of G11.2$-$0.3.  The greyscale shows difference images for $\alpha_t =
0.50$ to $0.60$ in steps of 0.02.  Contours correspond to 6-cm data
convolved with a 10$''$ circular beam (shown in lower right corners),
at levels of 5\% to 41\% of peak flux 91 mJy/beam, in steps of
12\%. \label{fig:snrtomog} }
\end{figure}
\par
Figure~\ref{fig:pwntomog} shows a series of tomographic images between
$0.15 < \alpha_t < 0.40$ in $\Delta \alpha = 0.05$ intervals.  The PWN
region in the center disappears into the surrounding region within the
shell when the trial spectral index is between $\alpha_t \approx$ 0.20
and 0.30.  Variations in the appearance of the shell interior between
one image and the next suggest that the rms noise here roughly
corresponds to a spectral index difference of $\sim 0.05$, although we
observe a very clear departure from uniform emission between 0.15 and
0.30; consequently, we choose asymmetric uncertainties around our best
estimate of $\alpha_{P} = 0.25^{+0.05}_{-0.10}$. 
\par
In Figure~\ref{fig:snrtomog}, we show tomographic images between $0.50
< \alpha_t < 0.60$ in $\Delta \alpha = 0.02$ intervals.  The majority
of the SNR shell fades into the surrounding background between
$\alpha_t \approx 0.54$ and $0.58$; we therefore estimate a value of
$\alpha_{S} = 0.56 \pm 0.02$ for most of the shell region.  However,
it appears that the SNR spectral index is not completely uniform, most
noticeably in the southeastern region.  At $\alpha_t = 0.52$, dark
patches can be seen in the outer shell; likewise, lingering positive
residuals appear as lighter emission in the inner shell of the same
region at $\alpha_t = 0.58$.  This gradient indicates that the
spectrum for the outer shell is slightly flatter, and the inner
shell slightly steeper, than for the bulk of the remnant.  However, it
should be noted that the similar size of the 6-cm beam relative to
G11.2$-$0.3 may have produced this apparent gradient if the primary
beam is not perfectly modeled.  Mosaiced \VLA observations are
currently underway which will verify this feature.
\par
Our measured values for $\alpha_{P}$ and $\alpha_{S}$  are in
agreement with single-dish data as published by \citet{koth01}, who
predict a fitted spectral index of 
$\alpha_{total} = 0.50 \pm 0.02$ from integrated flux density values
of the total remnant.  Assuming a flat spectrum core of
$\alpha_{P} = 0$, they subtracted the pulsar wind component of the
emission from the total, and inferred $\alpha_{S} = 0.57$.  This
value is well within the uncertainty of our result.  They also discuss
shell structure, specifically, the southeastern region's appearance of
being in a later evolutionary phase than the rest of the SNR, based on
polarization properties and X-ray observations.  The possibility of a
connection between  their conclusion and our detected gradient in
$\alpha$ should perhaps be investigated further.  

\subsection{Flux Density}
Using \textsc{miriad}, we determined the integrated flux over various
regions in the source, and subtracted the background signal estimated
from nearby sky.  We estimate the background-subtracted flux density
of the entire source to be $S_T = 16.6 \pm 0.9$ Jy at 20 cm and $8.4
\pm 0.9$ Jy at 6 cm.  Similarly, we subtract the background interior
to the shell surrounding the plerionic region to find the PWN flux
density $S_{P} = 0.36 \pm 0.23$ Jy (20 cm) and $0.32 \pm 0.18$ Jy
(6 cm).  The uncertainties are inferred from the rms variations of flux
in the background; however, while the exterior background was
relatively uniform in brightness, the interior background was highly
irregular, causing the large uncertainties in the PWN flux.  We then
estimate the SNR shell flux density to be $S_{S} = 16.2 \pm 1.1$ Jy
(20 cm) and $8.0 \pm 1.1$ Jy (6 cm) based on these measured results.  
\par
The greatest source of systematic uncertainty in these values is a
result of flux missing due to the nature of interferometric
measurements having incomplete $u-v$ coverage.  Single-dish
measurements would not be subject to the same uncertainty;
consequently, in comparing our results with those of \citet{koth01},
we estimate that $\sim 5$\% of the total 20-cm flux and $\sim 12$\% of
6-cm total flux is missing from our maps.  We also measure the amount
of missing 6-cm emission resulting from incomplete coverage by
comparing the flux levels found in the 20-cm image before and after
the $u-v$ modelling process.  This $\sim 11$\% difference in the total
object and $\sim 13$\% difference in the plerion indicates how much
flux is lost in the 20-cm data by degrading it to 6-cm visibility,
thereby implying that the 6-cm data are also missing $\sim
11$\%$-13$\% flux.

\section{DISCUSSION}

\subsection{Spectral Properties}
\label{sec:spec}
\begin{figure}[!h]
\plotone{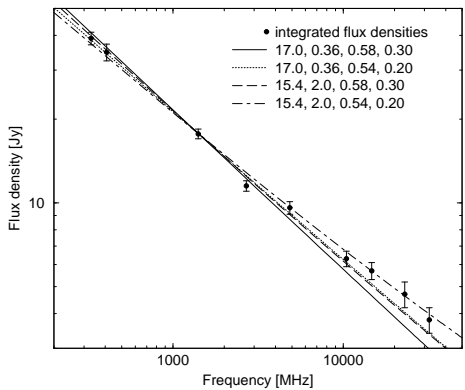}
\figcaption{Log-log plot of G11.2$-$0.3 flux density spectrum.  See
Table~\ref{tab:flux} for flux data points.  The fits correspond to 
the power-law spectrum (Eq.~\ref{equ:totflux}), with the upper
and lower bound parameters for $S_{S},\ S_{P},\ \alpha_{S}$, and
$\alpha_{P}$, respectively, as listed in the legend.  A detailed
description of the upper and lower limits is found in
\S~\ref{sec:spec}. \label{fig:spec} }
\end{figure}

\begin{deluxetable}{ccl}
\tablewidth{350pt}
\tablecaption{Integrated flux density measurements of G11.2$-$0.3
\label{tab:flux} } 
\tablehead{
\colhead{Frequency} & \colhead{Flux Density} & \colhead{Reference} \\
\colhead{(MHz)} & \colhead{(Jy)} & \colhead{} } 
\startdata
330   & 39.0 $\pm$ 2.0 & \citet{kass92} \\
408   & 34.8 $\pm$ 2.4 & \citet{kass89,shav70} \\
1408  & 17.7 $\pm$ 0.7 & \citet{reic90} \\
2695  & 11.5 $\pm$ 0.5 & \citet{reic84} \\
4850  &  9.6 $\pm$ 0.5 & \citet{koth01} \\
10450 &  6.3 $\pm$ 0.4 & \citet{koth01} \\
14700 &  5.7 $\pm$ 0.4 & \citet{koth01} \\
23000 &  4.7 $\pm$ 0.5 & \citet{koth01} \\
32000 &  3.8 $\pm$ 0.4 & \citet{koth01} \\
\enddata
\end{deluxetable}
We now compare the results of past studies (see Table~\ref{tab:flux})
with our measurements of spectral index and flux density at 20 cm.  To
do this, we plot an integrated fitted spectrum of the form  
\begin{equation}
\label{equ:totflux}
S_T = S_{S}(\nu/1461)^{-\alpha_{S}} + S_{P}(\nu/1461)^{-\alpha_{P}}\ ,
\end{equation}
where $\nu$ is the frequency in MHz.  Because of uncertainties in
background subtraction and incomplete $u-v$ coverage, we choose to
plot an upper and a lower estimate for each value of flux density.  We
set $S_{P} = 0.36 \pm 0.23$ as our low value estimate, because we used
a region of high flux as background (i.e. inside the shell), and also
because of missing interferometric flux.  By assuming the lowest
possible background (i.e. outside the shell), we find a high value for
$S_{P}$, which gives a low limit estimate for $S_{S}$ of $15.4 \pm
0.9$ Jy.  In order to account for possible missing interferometric
flux, we deduce high flux estimates by subtracting our lower bounded
values from interpolated total power results of \citet{koth01}.  The
high estimates for $S_{S}$ and $S_{P}$ are approximately $17.0 \pm
0.9$ Jy and $2.0 \pm 1.6$ Jy, respectively.  Figure~\ref{fig:spec}
shows Equation~\ref{equ:totflux} plotted with our best high and low
estimates of $S$, and maximum and minimum uncertainties of $\alpha$.
We find our values to be consistent with single-dish values, within
estimated uncertainties.  The higher frequency data points suggest
that the true value of $S_{P}$ may be close to our upper estimate; we
therefore assume $S_{P} \sim 1$ Jy for our rough calculations of radio
efficiency.

\subsection{Morphology and Age}
\begin{figure}[!h]
\plotone{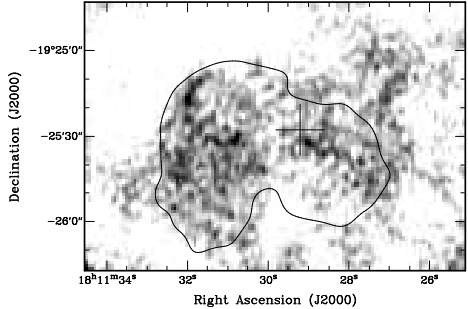}
\caption{Image of the PWN region in G11.2$-$0.3 at 20 cm.  The
contour represents roughly the PWN region boundary, as indicated by
the spectral tomography difference images.  The cross indicates the
location of the pulsar, as in Figure~\ref{fig:snrs}. \label{fig:plerion} }
\end{figure}

The tomography difference images allow us to identify each feature
according to its spectral index.  In particular, we recognize the
region of central emission that belongs to the PWN as outlined in the
$\alpha_{S}$ difference maps (Fig.~\ref{fig:snrtomog}) by the
prominent area of dark emission.  A contour of this region is
projected onto a detailed map of the plerionic region of G11.2$-$0.3
in Figure~\ref{fig:plerion}.  In this image, it is apparent that the
protruding feature in the northwestern direction, and possibly a
smaller, similar feature in the southeastern direction, is 
not spectrally part of the PWN, but may still be associated with
emission from the interior of the remnant.  A toroidal/jet morphology
has been detected in the Crab PWN, so it is possible that we may also
be seeing a similar shape here.  If it is a torus, then it is most
likely centered very near the pulsar position, with an axis of $\sim
30 \degr$ west of north.  There is some unexplained emission south
of its eastern branch, which appears to be associated with the PWN
but does not fit into the toroidal morphology, in addition to the
protrusions to the NW and SE, which are unassociated with the PWN.
Further discussion on the plerionic structure will be presented by
Roberts et al. (in preparation), as will the relationship between hard
X-ray and radio emission.
\par
We measure the angular diameter of the SNR to be $\sim 4\farcm5$ at
maximum extent, which corresponds to a linear diameter of $D_{S}
\sim 6.5$ pc, assuming the distance $d \sim 5$ kpc to the remnant is
accurate \citep{gree88}.  To calculate the average diameter of the
PWN, we find the total area encompassed by the plerion and approximate
it as radially symmetric, resulting in an angular diameter of $\sim
1\farcm3$ and $D_{P} \sim 1.8$ pc.  The \textit{Chandra} X-ray image
shows a PWN diameter of nearly $1\farcm3$ \citep{kasp01}, consistent
with radio results.  The ratio of diameters is $\sim 0.28$ in radio;
we compare this value against results for other composite remants and
find it to be higher than most other reported ratios \citep{vand01}.
Current theory predicts two main phases of PWN evolution inside the 
shell.  In the first phase, a bubble of plerionic gas freely expands
into the supernova ejecta supersonically, until it encounters the
reverse shock of the SNR blast wave, which compresses the plerion.  In
the second phase, the PWN continues to expand within the SNR at
subsonic velocities, as described by a Sedov solution
\citep{reyn84,vand01}.  Results of hydrodynamical numerical
simulations predict the largest radial extent ratio at the point just
before the reverse shock has reached the PWN (Fig.~8 of van der Swaluw
et al. 2000).  We detect a large PWN radius relative to its SNR;
therefore, this object has probably not yet encountered the reverse
shock.  Assuming these models are correct, this provides more evidence
that G11.2$-$0.3 is indeed a very young remnant, much younger than is
indicated by the 24,000 yr characteristic age of the pulsar
\citep{tori99,kasp01}.

\subsection{Efficiency} 
\citet{gaen00} characterize the efficiency $\epsilon$ of PWN radio
emission by the fractional contribution of spin-down energy to radio
luminosity,
\[ \epsilon = \frac{L_R}{\dot{E}}\ ,\]
where $L_R$ is the radio luminosity of the PWN and $\dot{E}$ is the
spin-down luminosity of the associated pulsar.  They
establish upper limits on $\epsilon$ by re-examining results
predicted by \citet{frai97}, with more up-to-date estimates for
interstellar medium densities.  Six of the eight other known radio PWN
are reported to have a ``typical'' efficiency of $\epsilon \sim
10^{-4}$.  Using our result of $\alpha_{P} \approx 0.25$ and
integrating over the range of 10 MHz to 100 GHz, we find
\[ \epsilon \sim 2 \times 10^{-4} \left ( \frac{S_{P}}{1\
\mathrm{Jy}} \right ) \left ( \frac{d}{5\ \mathrm{kpc}} \right )^2\ .\] 
Thus, the efficiency of G11.2$-$0.3 is consistent with those of other
detected PWN.

\section{CONCLUSIONS}
We have imaged SNR G11.2$-$0.3 at high resolution and revealed
the spectral properties of its individual components.  Our spectral
index and flux density results are consistent with recent studies of
this source and other composite-type SNR.  However, measurements
of its morphology suggest that it is in an earlier phase of evolution
(just entering the Sedov phase) than most other detected PWN inside
remnant shells, according to numerical models.  This result 
provides more evidence linking G11.2$-$0.3 to the supernova event of
386~\textsc{ad}.  We also find the radio efficiency to be as expected
for an object surrounding a young, energetic pulsar.

\acknowledgments
We wish to acknowledge Bryan Gaensler for helpful discussions and
David Green for sharing archival FITS files of G11.2$-$0.3.  This work
was supported in part by NSERC research grant RGPIN228738-00, an NSF
CAREER grant, and \textit{Chandra} grant GO0-1132X from the
Smithsonian Astrophysical Observatory to VMK.  The National Radio
Astronomy Observatory is a facility of the National Science Foundation
operated under cooperative agreement by Associated Universities, Inc.


\begin{thebibliography}{}

\bibitem[Clark \& Stephenson(1977)]{clar77} Clark, D.~H.~\&
Stephenson, F.~R.\ 1977, Oxford [Eng.] ; New York : Pergamon Press,
1977.~1st ed.
\bibitem[Crawford et al.(2001)]{craw01} Crawford, F., Gaensler, B.~M.,
Kaspi, V.~M., Manchester, R.~N., Camilo, F., Lyne, A.~G., \&
Pivovaroff, M.~J.\ 2001, \apj, 554, 152  
\bibitem[Downes(1984)]{down84} Downes, A.\ 1984, \mnras, 210, 845 
\bibitem[Frail \& Scharringhausen(1997)]{frai97} Frail, D.~A.~\&
Scharringhausen, B.~R.\ 1997, \apj, 480, 364
\bibitem[Gaensler et al.(1999)]{gaen99} Gaensler, B.~M., Brazier,
K.~T.~S., Manchester, R.~N., Johnston, S., \& Green, A.~J.\ 1999,
\mnras, 305, 724 
\bibitem[Gaensler et al.(2000)]{gaen00} Gaensler, B.~M., Stappers,
B.~W., Frail, D.~A., Moffett, D.~A., Johnston, S., \& Chatterjee, S.\
2000, \mnras, 318, 58 
\bibitem[Green et al.(1988)]{gree88} Green, D.~A., Gull, S.~F., Tan,
S.~M., \& Simon, A.~J.~B.\ 1988, \mnras, 231, 735
\bibitem[Kaspi et al.(2001)]{kasp01} Kaspi, V.~M., Roberts, M.~E.,
Vasisht, G., Gotthelf, E.~V., Pivovaroff, M., \& Kawai, N.\ 2001,
\apj, 560, 371
\bibitem[Kassim(1989)]{kass89} Kassim, N.~E.\ 1989, \apjs, 71, 799 
\bibitem[Kassim(1992)]{kass92} Kassim, N.~E.\ 1992, \aj, 103, 943 
\bibitem[Katz-Stone \& Rudnick(1997)]{katz97} Katz-Stone, D.~M.~\&
Rudnick, L.\ 1997, \apj, 488, 146  
\bibitem[Kothes \& Reich(2001)]{koth01} Kothes, R.~\& Reich, W.\ 2001,
\aap, 372, 627
\bibitem[Morsi \& Reich(1987)]{mors87} Morsi, H.~W.~\& Reich, W.\
1987, \aaps, 71, 189  
\bibitem[Narayan \& Nityananda(1986)]{nara86} Narayan, R.~\& 
Nityananda, R.\ 1986, \araa, 24, 127 
\bibitem[Reich et al.(1984)]{reic84} Reich, W., Fuerst, E., Haslam,
C.~G.~T., Steffen, P., \& Reif, K.\ 1984, \aaps, 58, 197
\bibitem[Reich, Reich \& Fuerst(1990)]{reic90} Reich, W., Reich, P.,
\& Fuerst, E.\ 1990, \aaps, 83, 539
\bibitem[Reynolds \& Chevalier(1984)]{reyn84} Reynolds, S.~P.~\&
Chevalier, R.~A.\ 1984, \apj, 278, 630
\bibitem[Reynolds et al.(1994)]{reyn94} Reynolds, S.~P., Lyutikov, M., 
Blandford, R.~D., \& Seward, F.~D.\ 1994, \mnras, 271, L1 
\bibitem[Sault \& Killeen(1999)]{saul99} Sault, R.~J.~\& Killeen,
N.~E.~B.\ 1999, The \textsc{miriad} User's Guide. Australia Telescope
National Facility, Sydney
(http://www.atnf.csiro.au/computing/software/miriad)
\bibitem[Shaver \& Goss(1970)]{shav70} Shaver, P.~A.~\& Goss, W.~M.\
1970, Australian Journal of Physics Astrophysical Supplement, 14, 77
\bibitem[Torii et al.(1997)]{tori97} Torii, K., Tsunemi, H., Dotani,
T.,~\& Mitsuda, K.\ 1997, \apjl, 489, L145
\bibitem[Torii et al.(1999)]{tori99} Torii, K., Tsunemi, H., Dotani,
T., Mitsuda, K., Kawai, N., Kinugasa, K., Saito, Y.,~\& Shibata, S.\
1999, \apjl, 523, L69 
\bibitem[van der Swaluw et al.(2000)]{vand00} van der Swaluw, E.,
Achterberg, A., Gallant, Y.~A.~\& T\'{o}th, G.\ 2000, \aap, submitted
(astro-ph/0012440)
\bibitem[van der Swaluw \& Wu(2001)]{vand01} van der Swaluw, E.~\& Wu,
Y.\ 2001, \apjl, 555, L49
\bibitem[Vasisht et al.(1996)]{vasi96} Vasisht, G., Aoki, T., Dotani,
T., Kulkarni, S.~R., \& Nagase, F.\ 1996, \apjl, 456, L59

\end{thebibliography}
\end{document}